\documentclass[11pt]{article}
\usepackage{graphicx}
\input epsf
\begin{document}
\begin{titlepage}
\begin{flushright}
CALT-68-2531\\
ITEP-TH-55/04
\end{flushright}

\begin{center}
{\large\bf $ $ \\ $ $ \\
Anomalous dimension and local charges}\\
\bigskip\bigskip\bigskip
{\large Andrei Mikhailov\footnote{e-mail: andrei@theory.caltech.edu}}
\\
\bigskip\bigskip
{\it California Institute of Technology 452-48,
Pasadena CA 91125 \\
\bigskip
and\\
\bigskip
Institute for Theoretical and 
Experimental Physics, \\
117259, Bol. Cheremushkinskaya, 25, 
Moscow, Russia}\\

\vskip 1cm
\end{center}

\begin{abstract}
AdS space is the universal covering of a hyperboloid.
We consider the action of the deck transformations on 
a classical string worldsheet in $AdS_5\times S^5$.
We argue that these transformations are generated by an infinite
linear combination of the local conserved charges. 
We conjecture that a similar relation holds for the 
corresponding operators on the field
theory side.
This would be a generalization of the recent field theory 
results showing
that the one loop anomalous dimension is proportional
to the Casimir operator in the representation of the
Yangian algebra.
\end{abstract}

\end{titlepage}

\section{Introduction}
The development of the AdS/CFT correspondence was originally obstructed
by our poor understanding of the string worldsheet theory in the background
with the nonzero Ramond-Ramond field strength. But recently it was realized
that the string worldsheet theory has a beautiful mathematical structure
related to the integrability. 
Classical integrability of the string worldsheet
theory first discussed in the context of the
AdS/CFT correspondence in \cite{MandalSuryanarayanaWadia,BPR}
should play an important role in the quantization of the worldsheet
theory. If we learn the proper use of these integrable
structures, this could perhaps even give us a fresh perspective on
the string perturbation theory.

Integrable structures were also found in the dual 
${\cal N}=4$ supersymmetric planar Yang-Mills theory.
Even before the discovery of the 
AdS/CFT correspondence, integrability was
found in the high energy sector of QCD in \cite{Lipatov,KorchemskyFaddeev};
see \cite{BBGK} for a recent discussion of the past results and
their relation to the string theory.
In \cite{MinahanZarembo,BeisertKristjansenStaudacher,SSC,BeisertStaudacher}
the one-loop anomalous dimension
of single trace operators in the ${\cal N}=4$ theory 
was computed; it was found
that the anomalous dimension is given by the Hamiltonian of 
an integrable spin chain. 
Moreover, some evidence of  the integrability at the level of two
loops was found;
see \cite{BeisertReview} for the review and the list of references.

Infinite dimensional symmetry algebras known as Yangians play 
an important role in the theory of integrable systems \cite{Bernard}. 
They are not Lie algebras, but rather associative algebras.
One of the most important properties of the Yangians is the
existence of the comultiplication which allows to introduce
a tensor product of representations. 
The comultiplication makes the Yangian a very natural algebra to consider
in the theory of spin chains. Indeed, the space of states of the spin chain
is the tensor products of the ``elementary'' spaces over its sites.
Usually Yangian acts in some elementary way on each site, and 
then we use the comultiplication to extend its action to the whole chain.
The action of the Yangian 
is encoded in the {\em transfer matrix} $T(z)$, which depends on
the {\em spectral parameter} $z$. The matrix elements of the
transfer matrix are not the c-numbers, but the linear
operators acting in the space of states of the spin chain. 
In other words, the Yangian algebra is generated by the
matrix elements of the transfer matrix $T(z)$; the coefficients
of the Taylor expansion of the matrix elements in powers of $z$
generate the Yangian. The defining relations of the Yangian
are encoded in the formula $R(z_1-z_2)T(z_1)T(z_2)=
T(z_2)T(z_1)R(z_1-z_2)$ where $R(z)$ is some matrix whose
elements are complex numbers. The form of this matrix is 
restricted by the conditions of the ``nontriviality''
of the algebra, known as the Yang-Baxter equations.

In the context of the AdS/CFT correspondence the Yangian symmetry was
first discussed on the string theory side in \cite{BPR}.
The study of the Yangian symmetry on the field theory side
was initiated in \cite{DNW1,DNW2}. It was conjectured that the
Yangian algebra acts on the single trace
states in the weakly coupled gauge theory. (Strictly speaking, one
has to consider operators constructed of infinitely many elementary fields,
in order to avoid some ``edge effects''. The action of the Yangian
requires a linear ordering of the sites of the chain, not just
a cyclic ordering.)
The explicit expression for the action in the zero coupling limit
was conjectured, motivated by the theory of spin chains.
In \cite{DolanNappi} the one-loop anomalous dimension was 
expressed in terms of the local conserved charges.
Local conserved charges can be obtained as the coefficients
of the Laurent series of $\log\mbox{tr}T(z)$ at the point
$z=z_0$ where $T(z)$ has a singularity\footnote{I want to thank
G.~Arutyunov for explaining this to me.}. They could be thought of as Casimir
operators of the Yangian. Even though the Yangian itself is not
well defined on the single trace states of a finite
engineering dimension, the Casimirs are well defined.

It would be interesting to extend the relation between the anomalous
dimension and the Casimirs to higher loops. 
The anomalous dimension is usually defined as the deformation of
the particular generator of the conformal algebra --- the
dilatation operator. But in fact the anomalous dimension parametrizes
the deformation of the {\em representation}
of the conformal algebra,
rather than the deformation of a particular generator. It is
more natural to define the anomalous dimension through the action
of the center of the conformal group. This definition is manifestly
conformally invariant. Notice that the coherent single-trace states
corresponding to the classical strings \cite{Kruczenski}
usually do not have a definite
engineering dimension \cite{SNS}, therefore the ``standard'' definition
will not work for them.

In  Section 2 we will explain how the anomalous dimension
can be defined through the action of the center of the conformal group.
In Section 3 we will show that this definition is very natural from
the string theory point of view and implies
that the anomalous dimension is an infinite sum of the
local conserved charges. 

\section{Anomalous dimension as a deck transformation.}
\subsection{Field theory side.}
The bosonic part of the supergroup $PSU(2,2|4)$ is not simply connected; 
the superconformal group of the conformal field theory is 
actually a covering group which we will denote 
$\widetilde{PSU}(2,2|4)$. 
The bosonic part of $\widetilde{PSU}(2,2|4)$
is $[\widetilde{SU}(2,2)\times SU(4)]/{\bf Z}_2$. We denoted
$\widetilde{SU}(2,2)$  the universal covering of
$SU(2,2)$, and ${\bf Z}_2$ is generated by 
$a\times (-{\bf 1})\in \widetilde{SU}(2,2)\times SU(4)$ where
$a$ is the rotation of the sphere $S^3$ by the angle $2\pi$
around one of its axes. (Notice that the bosonic part of
the superconformal group is not simply connected;
it has $\pi_1\simeq {\bf Z}_2$.)

Let $c$ denote the generator of the center.
The action of $c$ can be understood in the following way.
Consider the conformal field theory on ${\bf R}\times S^3$
where $\bf R$ is the time and the radius of $S^3$ is $1$.
Let $t$ denote the time and $\vec n$ denote the
unit vector parametrizing $S^3$. Then $c$ acts as the
combination of the conformal transformation:
\begin{equation}
	c:\;(t,\vec{n})\to (t+\pi,-\vec{n})
\end{equation}
with the R-symmetry $i{\bf 1}\in SU(4)$. 
This transformation
commutes with the generators of $so(2,4)$ and therefore
it is in the center of the conformal group. It also commutes
with the fermionic generators of the supersymmetry, therefore
it is in the center of $\widetilde{PSU}(2,2|4)$.

If we represent the elements of the group $U(2,2|4)$ by the
$(4|4)\times (4|4)$ matrices, then $c$ will correspond to 
the central element:
\begin{equation}\label{CentralMatrix}
	c=\left(
	\begin{array}{cc}
		i{\bf 1}_{4\times 4} & {\bf 0}_{4\times 4} \\
		{\bf 0}_{4\times 4} & i{\bf 1}_{4\times 4}
	\end{array}
	\right)
\end{equation}
But to describe  the superconformal group, we have to work on the covering
space of the space of matrices. 
Therefore, the matrix (\ref{CentralMatrix}) should be supplemented
with the path connecting it to the unit matrix, two choices of the
path considered equivalent if they can be smoothly deformed to each other.
The central element $c$ corresponds to the following path:
\begin{equation}
	C(t)=\left(
	\begin{array}{cc}
		\mbox{diag}\left(e^{ -{3\pi i\over 2}t}, 
		e^{ {\pi i\over 2}t}, e^{ {\pi i\over 2}t}, 
		e^{ {\pi i\over 2}t}\right) & {\bf 0}_{4\times 4}\\
		{\bf 0}_{4\times 4} & \mbox{diag}\left( 
		e^{ -{3\pi i\over 2}t}, 
		e^{ {\pi i\over 2}t}, e^{ {\pi i\over 2}t}, 
		e^{ {\pi i\over 2}t}\right) 
	\end{array}
	\right)
\end{equation}
parametrized by $t\in [0,1]$, $C(0)={\bf 1}$, $C(1)=c$.

Notice that 
$c^2$ acts as the shift of the time $t\to t+2\pi$ combined
with $(-1)^F$.
In the free field theory $c=1$ 
but in the interacting theory $c$ acts nontrivially. We will
define the anomalous dimension through the action of $c^2$:
\begin{equation}\label{DefinitionDeltaE}
	c^2=e^{2\pi i \Delta E}
\end{equation}
In perturbation theory
$\Delta E$ is expanded in powers of $\lambda$. We will call
$\Delta E$ the anomalous dimension. 
We could also define $\Delta E$ through the action of $c$,
but the corresponding definition on the string theory
side would be somewhat less transparent.

This definition is equivalent
to the definition throught the deformation of the dilatation 
generator for the operators which have a definite engineering dimension.
Indeed, the correction due to the interaction 
to the eigenvalue of the dilatation operator
on the local operator $\cal O$ in the Euclidean theory is equal
to the energy shift for the corresponding state $\psi$ in the theory on
${\bf R}\times S^3$. The energy is the eigenvalue of a generator
of the superconformal group which we will denote $H$.
Geometrically $H$ generates shifts of the global time ${\bf R}$.
Suppose that $\psi$ is an eigenstate of the Hamiltonian:
\begin{equation}
	H\psi=E\psi
\end{equation}
Let $E_0$ be the energy of this state in the free theory
(the engineering dimension of the corresponding operator).
Then in the interacting theory $E=E_0+\Delta E$ 
where $\Delta E$ is the energy shift.
In the Yang-Mills perturbation theory
$\Delta E\simeq \lambda$. We have
\begin{equation}
	e^{2\pi i H}\psi=e^{2\pi i E}\psi
	= e^{2\pi i (E_0+\Delta E)}\psi=(-1)^F e^{2\pi i\Delta E}\psi
\end{equation}
since $E_0$ is an integer or a half-integer depending on whether
the state is bosonic or fermionic.
In perturbation theory $\Delta E<< 1$ and therefore we can write
\begin{equation}\label{Explanation}
	\Delta E= {1\over 2\pi i} \log \left( (-1)^F e^{2\pi i H}\right)
\end{equation}
Notice that $e^{2\pi i H}$ generates the shift $t\to t+2\pi$.
The operator $(-1)^F e^{2\pi i H}$ is in the center of the superconformal
group (it commutes with all the generators of the superconformal group).
This operator  represents on the space of states
the discrete symmetry which we have denoted $c^2$. 
Therefore Eq. (\ref{Explanation}) implies Eq. (\ref{DefinitionDeltaE}).

\subsection{AdS side.}
The AdS space is the universal covering space of the
hyperboloid and $c^2$
acts as a deck transformation exchanging the sheets.
We can visualize the action of this deck transformation
on the string phase space in the following way.
Let us replace $AdS_5$  by the hyperboloid
$H=AdS_5/{\bf Z}$. Let us formally consider
the string as living on $(AdS_5/{\bf Z})\times S^5$.
Pick a point $x$ on the string worldsheet $\Sigma$.
Consider a neighborhood of $x$ in 
$(AdS_5/{\bf Z})\times S^5$ 
which is simply connected.
An example of such a neighborhood is a set
of points  which are within the distance
$R/2$ from $x$, where $R$ is the radius of $AdS_5$.
Let $B$ denote such a neighborhood. Consider
the part of the string worldsheet which is inside
$B$ (that is, $B\cap \Sigma$). One can see that
$B\cap \Sigma$ consists of several 
sheets, which can be enumerated. 
These sheets are two-dimensional, so we can think of
them as cards; $B\cap \Sigma$ is then a deck of 
cards, see Fig.1.
\begin{figure} 
\begin{center} 
\epsfxsize=2.4in {\epsfbox{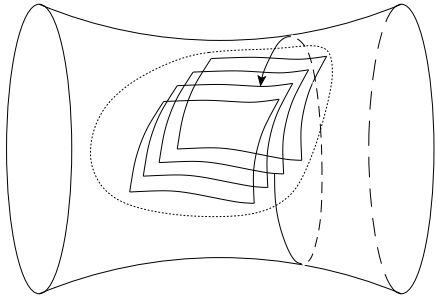}} 
\caption{The string worldsheet looks locally like a deck of cards.
Going around the noncontractible cycle in the hyperboloid
exchanges the sheets. This is the {\em deck transformation}.
It measures the deviation of the string worldsheet from being
periodic in the global time.}
\end{center} 
\end{figure}
Let $x$ belong to 
the sheet number $n$, then we can draw a path
on $\Sigma$ starting at $x$, winding once on 
the noncontractible cycle in $AdS_5/{\bf Z}$ and
then ending on the sheet number $n+1$. 
Let $\Sigma_n$ denote the sheet number $n$.
The deck transformation maps $\Sigma_n$ to $\Sigma_{n+1}$,
$\Sigma_{n+1}$ to $\Sigma_{n+2}$ and so on.
This determines the action of $c^2$ on the phase space of the
classical string.

\section{Deck transformations and local charges.}
Classical strings in AdS space correspond to classical
single-trace states on the field theory side.
In the classical regime, the length of the operator (the number
of the elementary fields under the trace) was conjectured to be
conserved. The corresponding conserved charge was constructed
in \cite{KT} using the Hamiltonian 
perturbation theory around the null-surfaces.
We have conjectured in \cite{Notes} that this conserved charge
is an infinite linear combination of the local conserved charges
which are known as Pohlmeyer charges \cite{Pohlmeyer}:
\begin{equation}\label{Expansion}
	L=\sqrt{\lambda}\left[{\cal E}_2+a_1{\cal E}_4+a_2{\cal E}_6+
	a_3{\cal E}_8+\ldots\right]
\end{equation}
 The coefficients $a_n$
can in principle be extracted\footnote{Note added in the revised version:
the coefficients $a_n$ are calculated 
from the plane wave limit in \cite{PWL}.}
from the known explicit expressions for the
charges of the rigid strings (see \cite{AS,Engquist} and the discussion
in \cite{Notes}) but it should be
possible to find a general interpretation of this conserved
charge in the framework of  the Bethe ansatz 
\cite{KMMZ,Marshakov,KazakovZarembo,BKS}. 
The characteristic property of
$L$ is that the corresponding Hamiltonian vector
field $\xi_L$ has periodic trajectories:
\begin{equation}\label{Periodic}
	e^{2\pi \xi_L}=\mbox{identical transformation}
\end{equation}
Therefore $L$ is an action variable.

The dynamics of the classical string
in $AdS_5\times S^5$ essentially splits into the direct product
of two systems: the sigma-model with the target space
$AdS_5$ and the sigma-model with the target space $S^5$.
 In Eq. (\ref{Expansion}) ${\cal E}_{2k}$
are the Pohlmeyer charges for the $S^5$ sigma-model.
One can construct in the same way the Pohlmeyer
charges for the $AdS_5$ sigma-model. Let us denote them ${\cal F}_{2k}$.
Consider the conserved charge $K$:
\begin{equation}\label{ExpansionAdS}
	K=\sqrt{\lambda}\left[{\cal F}_2+a_1{\cal F}_4+a_2{\cal F}_6+
	a_3{\cal F}_8+\ldots\right]
\end{equation}
defined with the same coefficients $a_k$ as in (\ref{Expansion}).
This charge would also generate the periodic trajectories if we
formally considered the string on $(AdS_5/{\bf Z})\times S^5$.
We would then have $e^{-2\pi\xi_K}$ acting as the identity map.
But for the string on $AdS_5\times S^5$ it acts as the 
deck transformation:
\begin{equation}
	e^{-2\pi\xi_K}=c^2
\end{equation}
Indeed, 
since $AdS$ is the universal cover of the hyperboloid,
the fact that the canonical transformation $e^{-2\pi\xi_K}$ acts as an 
identical map on the string on the hyperboloid implies that it acts
on the string on $AdS_5$ as some iteration
of the deck transformation. In fact it is the first iteration of
the deck transformation. To understand why it is  the first iteration
(and not, for example, the second iteration $c^4$) it is enough
to consider its action on the null-surface. 
The projection of the 
null-surface to the hyperboloid $AdS_5/{\bf Z}$ is a continuous collection of
equators of the hyperboloid.
Let us specify the null-surface by its embedding $x_0(\tau, \sigma)$;
for 
a fixed $\sigma=\sigma_0$ the curve $x_0(\tau,\sigma_0)$ is a 
light ray. We will denote
$x_{0,S}$ the projection of the string worldsheet to the sphere,
and $x_{0,A}$ the projection to AdS. We will use the worldsheet 
coordinates with the property 
$(\partial_{\tau}x_{0,A},\partial_{\tau}x_{0,A})=-1$.
 The vector field $\xi_K$ acts as 
an infinitesimal shift along the equator:
\begin{equation}
\begin{array}{l}
	\xi_K.\; x_{0,S}=0\\[5pt]
	\xi_K.\; x_{0,A}=\partial_{\tau} x_{0,A}
\end{array}
\end{equation}
Therefore $e^{-\alpha\xi_{K}}$ acts as a shift by an angle 
$\alpha$ along the equator. When $\alpha$ grows from $0$ to $2\pi$
the point $[e^{-\alpha\xi_{K}}x_0](\tau,\sigma)$ goes all the way
around the corresponding equator of AdS. As a point on the hyperboloid, it
returns when $\alpha=2\pi$ back to $x_0(\tau,\sigma)$; but as a point
on AdS it goes around the noncontractible cycle and ends up on the
other cover of the hyperboloid. This means that $e^{-2\pi \xi_{K}}$
acts as the first iteration of the deck transformation on the
null-surfaces. Since the null-surfaces can be approximated by
the fast moving strings, this implies by the continuity that 
$e^{-2\pi \xi_{K}}$ acts as $c^2$ on the string if the string
moves fast enough. 

Notice  that $\xi_K$ commutes with $\xi_L$.
Therefore 
\begin{equation}
	c^2=e^{2\pi(\xi_L-\xi_K)}
\end{equation}
The Hamiltonian flow of ${\cal E}_2-{\cal F}_2$ acts trivially (this
combination generates \hyphenation{re-pa-ra-metri-zations} 
reparametrizations on the string worldsheet;
the Virasoro constraints require that  ${\cal E}_2={\cal F}_2$).
Therefore we can identify
\begin{equation}\label{LogC}
	{1\over 2\pi}\log c^2= 
	\sqrt{\lambda}\left[a_1({\cal E}_4-{\cal F}_4)+
	a_2({\cal E}_6-{\cal F}_6)+a_3({\cal E}_8-{\cal F}_8)+\ldots
	\right]
\end{equation}
This expression is a perturbative expansion of the anomalous
dimension of the fast moving string in the null-surface perturbation
theory \cite{dVGN,SpeedingStrings}. 
The small parameter
is the relativistic factor $\sqrt{1-v^2}$, 
where $v$ is the typical velocity of the string.  
One can define the local conserved charges in such a way
that ${\cal E}_{2k}$ and ${\cal F}_{2k}$ are of the order
$(1-v^2)^{k-3/2}$  and  depend on the embedding coordinates
$x(\tau,\sigma)$ and their derivatives with respect
to $\tau$ and $\sigma$ up to the order $k$.
Eq. (\ref{Expansion}) and the fact that 
${\cal E}_{2}\sim (1-v^2)^{-1/2}$ imply that 
$\sqrt{1-v^2}\sim {\sqrt{\lambda}\over L}$ and
therefore (\ref{LogC}) is an expansion in powers
of $\lambda\over L^2$.
The first term is of the order $\lambda\over L$, the
second term is of the order $\lambda^2\over L^3$ and so on 
\cite{FT02}.
The zeroth approximation is the
null-surface corresponding to the infinitely long operator
which has a zero anomalous dimension.

It is not clear to us if the expansion (\ref{LogC})
converges and defines the action of the deck transformations
beyond the perturbation theory. The local conserved charges
do not exhaust all the commuting Hamiltonians of the sigma-model,
but the other conserved charges are nonlocal. In the
perturbation theory 
we know from \cite{KT,Notes} that $L$ is a local functional
in each order of the perturbation theory. Therefore the local charges
should be enough to construct $L$ in the perturbation theory.

On the field theory side ${\cal E}_{2k}-{\cal F}_{2k}$
should be identified with some combinations of
the Casimir operators of the Yangian (more precisely,
the purely bosonic
parts of the Casimir operators of the super-Yangian of $PSU(2,2|4)$).
It would be very interesting to extend the calculation
of \cite{DolanNappi} to higher loops and write the formula analogous
to (\ref{LogC}) on the field theory side. 

\section*{Acknowledgments}
I would like to thank  G.~Arutyunov, A.~Gorsky, J.~Polchinski and
A.~Tseytlin for discussions.
This research was supported by the Sherman Fairchild 
Fellowship and in part
by the RFBR Grant No.  03-02-17373 and in part by the 
Russian Grant for the support of the scientific schools
No. 00-15-96557.

\end{document}